# Decoupling of Photon Propagator in Compact QED

Ken Yee*

Dept. of Physics and Astronomy, LSU, Baton Rouge, Louisiana  70803-4001  USA

In compact QED$_{2+1}$ quantum monopole fluctuations induce confinement by expelling electric flux in a dual Meissner effect. Guided by Landau-Ginzburg theory, one might guess that the inverse London penetration depth $\lambda^{-1}$—the only physical mass scale—equals the photon propagator mass pole $M_\gamma$. I show this is not true. Indeed, in the Villain approximation the monopole part of the partition function factorizes from the photon part, whose dynamical variables are Dirac strings. Since Dirac strings are gauge-variant structures, I conclude that $M_\gamma$ is physically irrelevant: it is not a blood relative of $\lambda$ or any other quantity in the gauge-invariant sector. This result is confirmed by numerical simulations in the full theory, where $M_\gamma$ is not sensitive to monopole prohibition but essentially vanishes if Dirac strings are prohibited.

## 1. The Issue

In the semiclassical superconducting model of the QCD vacuum [1], in which electric flux is restricted to Abrikosov tubes of width $\lambda$, the inverse London penetration depth $\lambda^{-1}$ is the mass of an effective gauge potential $A_\mu^{\rm eff}$. Recent calculations of $\lambda$ [2] and, independently, of gluon $A_\mu$ propagator mass poles $M$ in LGT(lattice gauge theory) [3] present the question: How, if at all, is $A_\mu^{\rm eff}$ related to $A_\mu$ and $\lambda$ to $M$? In this talk I demonstrate in cQED$_{2+1}$(compact QED in $2+1$ dimensions), a QCD-like LGT, that the physically relevant quantities $A_\mu^{\rm eff}$ and $\lambda$ are unrelated to $A_\mu$ and $M$, which are unphysical.

## 2. Monopoles and Dirac Strings

In *non*compact QED, the photon is an unbounded real field $a_\mu \in (-\infty,\infty)$ and the action[2] $S_0 = \frac{\beta}{4}\sum_{\mu,\nu} f_{\mu\nu}^2$ where $f_{\mu\nu} \equiv \partial_\mu a_\nu - \partial_\nu a_\mu$ is gauge-invariant under $\delta a_\mu = -\partial_\mu \omega_x$. Since $S_0$ is gaussian, the nonperturbative photon mass $M_\gamma$ vanishes because Maxwell equation $\partial_\mu^* f_{\mu\nu} = 0$ in Landau gauge implies $\Box a_\mu = 0$.

Nothing is wrong with noncompact QED except that $S_0$ does not have natural nonabelian extensions. The $U(1)$ LGT corresponding to lattice QCD is cQED. Links $U_\mu \equiv e^{-i\theta_\mu}$ in cQED depend only on the photon $A_\mu$ part of

$$\theta_\mu \equiv A_\mu + 2\pi n_\mu \quad -\pi \leq A_\mu < \pi \ . \quad (1)$$

$A_\mu$ is the lattice photon whose nonperturbative propagator mass $M_\gamma$ is of concern. cQED$_{2+1}$ has local gauge invariance, chiral symmetry breaking, and area-law electron confinement induced by quantum monopole percolation [1]. cQED photons are uncharged but they suffer confinement since, heuristically, the "adjoint" Wilson loop obeys $\langle \prod_{l \in \text{ loop}} \sin\theta_l \rangle \propto \text{Re} \langle \prod_{l \in \text{ loop}} e^{i\theta_l} \rangle$ where cross terms are suppressed by gauge invariance. Therefore electron confinement implies photon confinement and cQED photons, like QCD gluons, are confined.

The cQED action is $S_c \equiv \beta \sum_{\mu<\nu}\left(1-\cos F_{\mu\nu}\right)$ where the plaquette angle is

$$F_{\mu\nu} \equiv \partial_\mu A_\nu - \partial_\nu A_\mu. \quad (2)$$

$S_c$ is invariant under local gauge transformation

$$\delta\theta_\mu \equiv -\partial_\mu \omega_x, \quad (3)$$

$$\delta A_\mu \equiv (A_\mu - \partial_\mu\omega_x)\text{Mod}[-\pi,\pi] - A_\mu. \quad (4)$$

While plaquette $\exp(iF_{\mu\nu})$ is gauge-invariant, a gauge transformation inducing unequal shifts of $n_\mu$ on the four links of $F_{\mu\nu}$ shifts $F_{\mu\nu}$ by a $2\pi$ multiple. $F_{\mu\nu}$ decomposes into a gauge-invariant physical part $\Theta_{\mu\nu} \in [-\pi,\pi)$ and a gauge-variant integral kink $N_{\mu\nu} \in \mathbf{Z}$ such that

$$F_{\mu\nu} \equiv (\Theta + 2\pi N)_{\mu\nu}, \quad (5)$$

$$\delta F_{\mu\nu} = 2\pi\delta N_{\mu\nu} = \partial_\mu(\delta\theta - \delta A)_\nu - (\mu \leftrightarrow \nu). \quad (6)$$

---

*Email: kyee@rouge.phys.lsu.edu
[2]$\partial_\mu h_x \equiv h_{x+\hat\mu} - h_x$, $\partial_\mu^* h_x \equiv h_x - h_{x-\hat\mu}$, and $\Box \equiv \partial_\mu^*\partial_\mu$. I will ignore topological gauge transformations.

The key feature of $S_c$ is $\cos F_{\mu\nu} = \cos \Theta_{\mu\nu}$, required by gauge invariance as $N_{\mu\nu}$ is locally gauge-variant. While any single $N_{\mu\nu}$ can be gauged away, spatial combinations of them form gauge-invariant structures which influence $\Theta_{\mu\nu}$. To see how this works, decompose according to the Hodge-DeRham theorem

$$\Theta_{\mu\nu} = \epsilon_{\mu\nu\alpha} \partial_\alpha^* \phi + \partial_\mu \alpha_\nu - \partial_\nu \alpha_\mu, \tag{7}$$

$$N_{\mu\nu} = \epsilon_{\mu\nu\alpha} \partial_\alpha^* m + \partial_\mu l_\nu - \partial_\nu l_\mu \tag{8}$$

where $\phi, \alpha_\mu \in (-\infty, \infty)$ and $m, l_\mu \in \mathbf{Z}$. $\phi$ and $m$ are invariant under (3), $\alpha_\mu$ transforms like $\theta_\mu$, and $l_\mu$ like $(A-\alpha)_\mu/2\pi$. $\Theta_{\mu\nu}$ (and similarly $N_{\mu\nu}$) has 3 independent polarizations while $\phi$ and $\alpha_\mu$ are 4 functions because $\Theta_{\mu\nu}$ is invariant under $\delta\alpha_\mu = -\partial_\mu \omega_x$.

In vector notation Eq. (5) becomes

$$\vec{B} = \vec{H} + 2\pi \vec{\eta} \tag{9}$$

where the total $\vec{B}$ and physical $\vec{H}$ magnetic(actually electromagnetic) fields are

$$\vec{B} \equiv \nabla \times \vec{A}, \quad \vec{H} \equiv \nabla \phi + \nabla \times \vec{\alpha}. \tag{10}$$

It will be advantageous to recast Dirac string field

$$\vec{\eta} = \nabla m + \nabla \times \vec{l} \tag{11}$$

in terms of its divergence and curl

$$q \equiv \nabla \cdot \vec{\eta} = \Box m, \tag{12}$$

$$\vec{\rho} \equiv \nabla \times \vec{\eta} = \nabla(\nabla \cdot \vec{l}) - \Box \vec{l}. \tag{13}$$

Since $\nabla \cdot \vec{B} = 0$ by (10), $\nabla \cdot \vec{H} = -2\pi q$, that is, $q$ causes dislocations in the physical field $\vec{H}$. For example, let $s(t)$ be the step function. A monopole at the origin attached to a string along the positive $\hat{t}$-axis corresponds to $\eta_\mu = \delta_{\mu,0} \delta_{x,0} \delta_{y,0} s(t)$, $q_x = \delta_{x,0} \delta_{y,0} \delta_{t,0}$ and $\vec{\rho} = (\delta'_{x,0} \delta_{y,0} \hat{y} - \delta_{x,0} \delta'_{y,0} \hat{x}) s(t)$. By tautology, $q$ is the magnetic monopole density, gauge invariant since $m$ is gauge invariant. In contrast $\vec{\rho}$, a continuous current wrapping around $\vec{\eta}$, is gauge-variant.

In general, kinks occur either in monopoles, Dirac strings connecting a monopole antimonopole pair, or Dirac string loops. Loops can either be homologically trivial or toroidally wind around the periodic boundaries.[3] Monopole charge density $q$ is gauge-invariant but the number of string loops and the length and shape of all strings vary with gauge. Segments of string $\vec{\eta}$ are characterized by $\vec{\rho} = \nabla \times \vec{\eta}$, continuous flows winding around $\vec{\eta}$.

### 3. Difference Between $\vec{A}^{\text{eff}}$ and $\vec{A}$

Upon adopting a condition such as $\nabla \cdot \vec{l} = 0$ and ignoring Laplacian zero modes, Eqs. (5)-(13) constitute 1-to-1 variables changes

$$\{N\} \to \{m, \vec{l}\} \to \{q, \vec{\rho}\} \to \{\phi, \vec{\alpha}\} \tag{14}$$

where, if $\Box \Delta_{x,y} = -\delta_{x,y}$, $\partial_\mu^* \Delta^{\mu\nu} = 0$, and $\Box \Delta_{x,y}^{\mu\nu} = -\delta_{\mu,\nu} \delta_{x,y}$, then

$$\phi = 2\pi \int_y \Delta_{x,y} q_y, \quad \alpha_\mu = \vec{A}_\mu - 2\pi \int_y \Delta_{x,y}^{\mu\nu} \vec{\rho}_{\nu y}. \tag{15}$$

In Villain's periodic gaussian approximation[4] $S_c \to S_c^V$ where following (5) and (7)

$$Z_c \equiv \int_A e^{-S_c^V} \equiv \sum_{\{N\}} \int_A e^{-\frac{\beta}{4} \sum_x (F[A] - 2\pi N)^2} \tag{16}$$

$$= \sum_{\{q,\vec{l}\}} \int_A e^{-\frac{\beta}{4} \sum_x F^2[A - 2\pi l] + 2(\nabla \phi)^2} \tag{17}$$

$$= Z_m[0] \times Z_{Al}[0], \tag{18}$$

$$Z_m[\xi] \equiv \sum_{\{q\}} e^{\sum_x \xi_x q_x - 2\pi^2 \beta \sum_{x,y} q_x \Delta(x-y) q_y}, \tag{19}$$

$$Z_{Al}[0] = \int_\alpha e^{-\frac{\beta}{4} \sum_x F^2[\alpha]}, \quad \alpha_\mu \in (-\infty, \infty). \tag{20}$$

The sum over $\{N\}$ in (16) maintains gauge invariance under (6). (18) follows from

$$\sum_x \epsilon_{\mu\nu\lambda} F_{\mu\nu} \partial_\lambda^* \phi = -\sum_x \phi \, \epsilon_{\mu\nu\lambda} \partial_\lambda F_{\mu\nu} = 0, \tag{21}$$

(19) from (15), and (20), which says $\vec{\alpha}$ is a massless noncompact photon, from calculus identity

---

[3] While my numerical gauge configurations have many string loops, I have found no homologically nontrivial ones.
[4] I will employ shorthand such as $F_{\mu\nu}[\alpha] \equiv \partial_\mu \alpha_\nu - (\mu \leftrightarrow \nu)$, $F \cdot G \equiv \sum_{\mu,\nu} F_{\mu\nu} G_{\mu\nu}$ and $\sum_{\{N\}} \equiv \prod_{\mu,\nu} \sum_{N_{\mu\nu} = -\infty}^\infty \delta(-N_{\nu\mu}, N_{\mu\nu})$.

$\sum_{l=-\infty}^{\infty} \int_{A=-\pi}^{\pi} h(A - 2\pi l) = \int_{\alpha=-\infty}^{\infty} h(\alpha)$.
Eq. (18) also relies on the quadratic character of the Villain approximation; keeping $\mathcal{O}(\Theta^4)$ terms in the $\cos \Theta_{\mu\nu}$'s of $S_c$ would destroy factorization.

Let us make contact with $A_\mu^{\text{eff}}$. Following Polyakov [1] the dilute gas expansion and occupation number resummation over $q \in \{0, \pm 1\}$ of $Z_m$ in (19) yields

$$Z_m[\xi] \propto \int_\Phi e^{-\frac{1}{4\pi^2\beta} \sum_x (\nabla(\Phi-\xi))^2 - 2\lambda^{-2}\cos\Phi} \quad (22)$$

where $\lambda^2 = 2\pi^2\beta e^{-2\pi^2\beta\Delta(0)}$. Dummy scalar $\Phi$ is semiclassically identified with $\phi$ in (7) via (15) because $\sum_x \xi_x q_x = \sum_{x,y} \nabla\xi \cdot \nabla \cdot \Delta_{x,y} q_y$. Comparing (19) to (22) implies for $\xi \to 0$

$$\sum_y \nabla \cdot \Delta_{x,y} \langle q_y \rangle_m = \frac{1}{Z_m} \frac{\delta Z_m}{\delta \nabla\xi} = \frac{\langle \nabla\Phi \rangle_\Phi}{2\pi^2\beta} \quad (23)$$

where $\langle \ \rangle_S$ refers to the expectation associated with partition function $Z_S$. Hence with $\vec{V} \equiv \nabla \times \vec{\alpha}$,

$$\langle \nabla \cdot \vec{H} \rangle_{Alm} = -2\pi \langle q \rangle_m = 0, \quad (24)$$

$$\langle \vec{H}_y \vec{H}_x \rangle_{Alm} = \langle \vec{V}_y \vec{V}_x \rangle_{Al} + \frac{4\pi^2}{Z_m} \frac{\delta^2 Z_m}{\delta\nabla\xi_y \delta\nabla\xi_x}. \quad (25)$$

If $\beta/\lambda \ll 1$, (24) and (25) are reproduced by an $M_\gamma = \lambda^{-1}$ *free* photon $\vec{A}^{\text{eff}}$ with $\vec{H}^{\text{eff}} \equiv \nabla \times \vec{A}^{\text{eff}}$, that is,

$$\nabla \cdot \vec{H}^{\text{eff}} = 0, \quad \langle \vec{H}_y^{\text{eff}} \vec{H}_x^{\text{eff}} \rangle_{\text{eff}} \approx \langle \vec{H}_y \vec{H}_x \rangle_{Alm}. \quad (26)$$

The second relation in (26) relies on masslessness of $\vec{\alpha}$ in (25), shown in (20), and $\cos\Phi \to -\Phi^2/2$ in (22). $\vec{A}^{\text{eff}}$ is the massive Landau-Ginzburg photon and $\lambda$ the London penetration depth.

The $\vec{A}$ propagator is generated by $Z_{Al}[J]$, defined by adding $J \cdot A$ to the action in (17), which does not affect factorization result (18). Thus the $\vec{A}$ mass $M_\gamma$ has nothing to do with monopoles $q$ and, hence, nothing to do with $\vec{A}^{\text{eff}}$ or $\lambda$. In contrast to the mass of $\vec{\alpha}$, $M_\gamma$ may be nonzero since $J \cdot A$ breaks the pure $\vec{\alpha}$-dependent form of $Z_{Al}[0]$. Manipulations like those leading to (19) yield $Z_{Al}[J] =$

$$\int_A \sum_{\{\vec{\rho}\}} e^{\sum_x (J+\pi\beta\rho) \cdot A - \frac{\beta}{4}F^2 - \pi^2\beta \sum_y \rho \cdot \Delta \cdot \rho}. \quad (27)$$

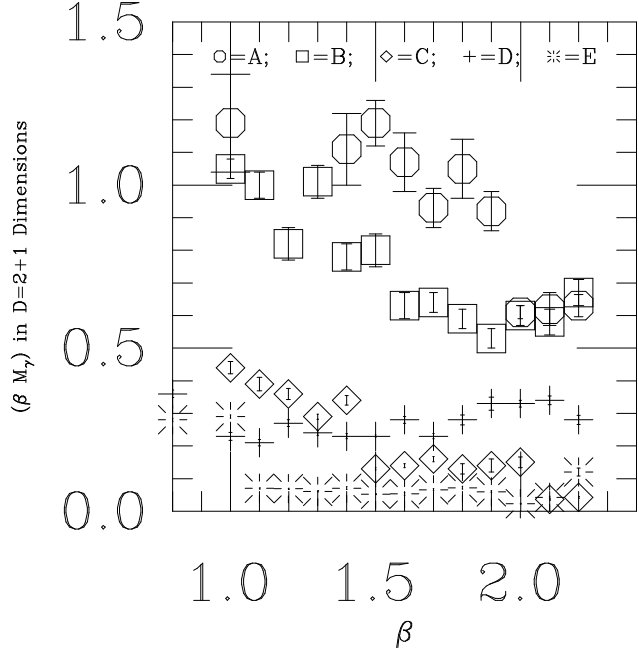

Figure 1. $\beta M_\gamma$ in five QED$_{2+1}$ variations.

Summing $\{\vec{\rho}\}$ is equivalent to summing Dirac string configurations. In Landau gauge $\nabla \cdot \vec{\rho} = 0$ and $Z_{Al}$ is the partition function of a Coulombic $\vec{\rho}$ loop gas. Interestingly $\vec{\rho}$ is a mixed state in the gas since for $M_\gamma$ to be nonzero

$$\int_{y,y'} \Delta_{x,y}^{\mu\alpha} \Delta_{0,y'}^{\nu\beta} \langle \rho_{y,\alpha} \, \rho_{y',\beta} \rangle_{\vec{\rho}} \quad (28)$$

must have a negative norm massless mode to cancel the $\alpha$ pole and an independent $M_\gamma$ mode.

In conclusion the $\vec{A}$ propagator decouples from monopoles $q$ in the Villain approximation and, accordingly, $M_\gamma$ is independent of the London penetration depth. Numerical experiments described in Section 4 support this result in full cQED.

## 4. Numerical Experiments

Figure 1 shows that Landau gauge $M_\gamma$ in cQED("A") is relatively insensitive to monopole prohibition("B") but dramatically reduced by kink prohibition("C" and "E"). Kinks are prohibited either by inserting a delta function in the link measure("C") or by replacing $\cos F_{\mu\nu}$ in $S_c$ with $\sim F_{\mu\nu}^2$("E"). The action for E is not invari-ant under kink-creating gauge transformations, which are also prohibited. Restoring the possibility of such gauge transformations during Landau gaugefixing("D") does not affect $M_\gamma$ much. This indicates that the kinks responsible for $M_\gamma$ in A are from the pre-gaugefixing configurations and not specifically created during Landau gaugefixing. (I suspect gaugefixing gives smaller string loops than those responsible for the bulk of $M_\gamma$.) At $\beta = 1.8$ the kink number density for cases A-E are $\rho_A = .41(.01)$, $\rho_C \equiv 0$, $\rho_D = .23(.004)$, and $\rho_E \sim 10^{-5}$. Since the $\beta = 1.8$ monopole number density is $8.0(1.1)10^{-3}$, forbidding monopoles doesn't change the kink density and $\rho_B = \rho_A$.

$\beta \times M_\gamma$ in the Figure, a dimensionless number in $D = 2 + 1$, is the log of the ratio of successive $\vec{p} = 0$ photon propagator timeslices. The central value of A is from 500 $S_c$-based configurations on $17^2 \times 19$ lattices. The first configuration is thermalized by 500 forty-hit, 40%-acceptance Metropolis sweeps and 5000 checkerboard gaugefixing sweeps. Configurations thereafter are separated by 5 forty-hit Metropolis sweeps and 5000 checkerboard gaugefixing sweeps. Errors are jackknife sigmas based on 10 450-configuration subaverages. Configurations $1-50$ are omitted from the first subaverage, $51-100$ from the second, $\cdots$. The numerical photon operator and gauge condition are $S_\mu \equiv \sin A_\mu$ and $\partial_\alpha^* S_\mu = 0$. $S_\mu$ corresponds to the gluon operator used in QCD simulations [3]. Since $\sin A_\mu = \sin(\pi - A_\mu)$, $S_\mu$ leaves $A_\mu$ ambiguous in reflections about $\pm \pi/2$.

B, Landau gauge cQED with monopoles $q$ prohibited, refers to configurations generated according to $S_c$ with the insertion of delta function $\prod_{\{x\}} \delta_{q,0}$ into the link measure. This is implemented starting with the $\theta_\mu = 0$ configuration and linkwise forbidding updates which create monopoles. Landau gaugefixing, which cannot change $q$, proceeds normally. C refers to $S_c$ configurations with the insertion of kink-forbidding delta function $\prod_{\{N\}} \delta_{N,0}$ into the measure. This insertion affects Landau gaugefixing by forbidding kink-creating gauge transformations. Due to this restriction, a good Landau gauge is not achieved but the photon propagator signal is strong. The tiny residual mass is due to $\mathcal{O}(\Theta_{\mu\nu}^4)$ terms in $S_c$ which ruin factorization (18).

D and E are based on the action $S_E = \frac{\beta}{4} \sum_x F^2$ where (2) defines $F_{\mu\nu}$. Unlike $S_c$, $S_E$ is invariant only under gauge transformations which preserve $N_{\mu\nu}$. E refers to $S_E$ configurations put as close as possible to Landau gauge with kink-changing gauge transformations forbidden. D refers to $S_E$ configurations fixed to Landau gauge by the *full set* of cQED gauge transformations. From the $S_E$ standpoint D, corrupted by action-changing kink creation and annihilation, is gauge *in*equivalent to E. The difference between $M_\gamma$ in D and E, gauge equivalent from the $S_c$ viewpoint, indicates how much kinks generated by the Landau gaugefixing algorithm contribute to $M_\gamma$.

## 5. Acknowledgements

I have benefitted from discussions on aspects of monopoles, superconductivity, and effective actions with Dana Browne, Vandana Singh, H.R. Fiebig, Lai-Him Chan, Claude Bernard, and especially Dick Haymaker, who inspired me to think about the London relation. KY is supported by DOE grant DE-FG05-91ER40617; computation was at the LSU Concurrent Computing Lab.